\pdfoutput=1

\documentclass[sigconf]{acmart}
\usepackage[utf8]{inputenc}
\usepackage[english]{babel}
\usepackage{array}
\usepackage{multirow}
\usepackage{graphicx} 
\usepackage{url}
\usepackage{subcaption}
\usepackage{color}
\usepackage{environ}
\usepackage{siunitx}
\usepackage{tikz}
\usetikzlibrary{fit, arrows, calc, positioning, backgrounds, shapes.misc, shapes.geometric}
\hypersetup{
    pdftitle={A TOCTOU Attack on DICE Attestation},
    pdfkeywords={IoT, Trusted computing, Attestation, DICE, TCG, ROP, Malware},
    bookmarksnumbered=true,     
    bookmarksopen=true,         
    bookmarksopenlevel=1,       
    colorlinks=false,            
    pdfstartview=Fit,           
    pdfpagemode=UseOutlines,    
    pdfpagelayout=TwoPageRight,
    urlcolor=[rgb]{0,0,0.5},
    linkcolor=[rgb]{0,0,0.5},
    citecolor=[rgb]{0,0,0.5}
}
\usepackage{hypcap}
\usepackage[]{algorithm2e}
\usepackage{bytefield}
\usepackage{mathtools}
\usepackage{placeins}
\usepackage{tabularx, ragged2e}
\usepackage[most]{tcolorbox}

\usepackage{enumitem}
\usepackage{booktabs}
\usepackage{float}
\usepackage{caption} 
\usepackage{newfloat}
\usepackage{booktabs}
\usepackage{glossaries}
\usepackage{cleveref}
\usepackage{adjustbox}
\usepackage{multicol, blindtext}
\usepackage{listings}
\usepackage{makecell}
\usepackage{chronology}

\usepackage{xcolor}
\usepackage{etoolbox}
\apptocmd{\sloppy}{\hbadness 10000\relax}{}{}

\definecolor{codegreen}{rgb}{0,0.6,0}
\definecolor{codegray}{rgb}{0.5,0.5,0.5}
\definecolor{codepurple}{rgb}{0.58,0,0.82}
\definecolor{backcolour}{rgb}{0.95,0.95,0.92}

\lstdefinestyle{mystyle}{
    language={[x86masm]Assembler},
    commentstyle=\color{codegreen},
    keywordstyle=\color{blue},
    numberstyle=\tiny\color{codegray},
    stringstyle=\color{codepurple},
    basicstyle=\ttfamily\footnotesize,
    breakatwhitespace=false,         
    breaklines=true,                 
    captionpos=b,                    
    keepspaces=true,                 
    numbersep=5pt,                  
    showspaces=false,                
    showstringspaces=false,
    showtabs=false,                  
    tabsize=2,
    frame=single,
    morekeywords={cpsid, for, do, end, ldr, to, i, movw, movt},
}
\title{A TOCTOU Attack on DICE Attestation}

\newacronym{dice}{DICE}{Device Identifier Composition Engine}
\newacronym{tcg}{TCG}{Trusted Computing Group}
\newacronym{rop}{ROP}{Return-Oriented Programming}
\newacronym{ble}{BLE}{Bluetooth Low Energy}
\newacronym{cdi}{CDI}{Compound Device Identifier}
\newacronym{uds}{UDS}{Unique Device Secret}
\newacronym{iot}{IoT}{Internet of Things}
\newacronym{soc}{SoC}{System on Chip}
\newacronym{tpm}{TPM}{Trusted Platform Module}
\newacronym{mac}{MAC}{Message Authentication Code}
\newacronym{mpu}{MPU}{Memory Protection Unit}
\newacronym{mmu}{MMU}{Memory Management Unit}
\newacronym{acl}{ACL}{Access Control List}
\newacronym{nvmc}{NVMC}{Non-Volatile Memory Controller}
\newacronym{sdk}{SDK}{Software Development Kit}
\newacronym{isa}{ISA}{Instruction Set Architecture}
\newacronym{mtu}{MTU}{Maximum Transmission Unit}
\newacronym{dos}{DoS}{Denial of Service}
\newacronym{wxr}{W$\oplus$X}{Write xor Execute}
\newacronym{os}{OS}{Operating System}
\newacronym{rtos}{RTOS}{Real-Time Operating System}
\newacronym{ddos}{DDoS}{Distributed Denial of Service} 
\newacronym{tee}{TEE}{Trusted Execution Environment}
\newacronym{awdt}{AWDT}{Authenticated Watchdog Timer}
\newacronym{aslr}{ASLR}{Address-Space Layout Randomization}
\newacronym{cpu}{CPU}{Central Processing Unit}
\newacronym{dep}{DEP}{Data Execution Prevention}
\newacronym{esp}{ESP}{Executable Space Protection}
\newacronym{ssp}{SSP}{Stack Smashing Protection}
\newacronym{toctou}{TOCTOU}{Time-Of-Check Time-Of-Use}
\newacronym{cfi}{CFI}{Control-Flow Integrity}

\newtcolorbox{protocolbox}{%
	enhanced, 
	unbreakable, 
	arc=0pt, 
	outer arc=0pt, 
	leftrule=0mm, 
	rightrule=0mm,
	colback=white,
	right=1mm,
	left=1mm
}

\newcounter{protocol}

\DeclareFloatingEnvironment[
	fileext=frm,
	name=Listing
]{listing}

\DeclareCaptionSubType{listing}

\copyrightyear{2022}
\acmYear{2022}
\setcopyright{acmlicensed}\acmConference[CODASPY '22]{Proceedings of the Twelveth ACM Conference on Data and Application Security and Privacy}{April 24--27, 2022}{Baltimore, MD, USA}
\acmBooktitle{Proceedings of the Twelveth ACM Conference on Data and Application Security and Privacy (CODASPY '22), April 24--27, 2022, Baltimore, MD, USA}
\acmPrice{15.00}
\acmDOI{10.1145/3508398.3511507}
\acmISBN{978-1-4503-9220-4/22/04}

\author{Stefan Hristozov}
\affiliation{%
  \institution{Fraunhofer AISEC}
  \city{Garching near Munich}
  \country{Germany}}
\email{stefan.hristozov@aisec.fraunhofer.de}

\author{Moritz Wettermann}
\affiliation{%
  \institution{Fraunhofer AISEC}
  \city{Garching near Munich}
  \country{Germany}}
\email{moritz.wettermann@aisec.fraunhofer.de}

\author{Manuel Huber}
\affiliation{%
  \institution{Microsoft}
  \city{Vancouver}
  \country{Canada}}
\email{manuel.huber@microsoft.com}

\ccsdesc[500]{Security and privacy~Embedded systems security}

\keywords{Attestation, DICE, TOCTOU, IoT, Trusted computing, TCG, ROP, Malware}

\begin{abstract}
A major security challenge for modern \gls{iot} deployments is to ensure that the devices run legitimate firmware free from malware. This challenge can be addressed through a security primitive called \textit{attestation} which allows a remote backend to verify the firmware integrity of the devices it manages. In order to accelerate broad attestation adoption in the \gls{iot} domain the \gls{tcg} has introduced the \gls{dice} series of specifications. \gls{dice} is a hardware-software architecture for constrained, e.g., microcontroller-based \gls{iot} devices where the firmware is divided into successively executed layers. 
 
In this paper, we demonstrate a remote \gls{toctou} attack on \gls{dice}-based attestation. We demonstrate that it is possible to install persistent malware in the flash memory of a constrained microcontroller that cannot be detected through \gls{dice}-based attestation. The main idea of our attack is to install malware during runtime of application logic in the top firmware layer. The malware reads the valid attestation key and stores it on the device's flash memory. After reboot, the malware uses the previously stored key for all subsequent attestations to the backend. We conduct the installation of malware and copying of the key through \gls{rop}. As a platform for our demonstration we use the Cortex-M-based nRF52840 microcontroller. We provide a discussion of several possible countermeasures which can mitigate  the shortcomings of the \gls{dice} specifications.

\end{abstract}

\begin{document}
\maketitle

\sloppy
\section{Introduction}
\gls{iot} deployments often incorporate many physically distributed, constrained (e.g., microcontroller-based) devices having minimal or even no security features due to cost reasons. At the same time, the devices have identical software stacks and configurations which allows scalable attacks to be developed once a vulnerability is discovered by an attacker. 
These properties make \gls{iot} deployments an attractive target for remote software attacks such as malware infections as demonstrated in \cite{mirai, Ronen2017, Weidler2017}.
Once malware is installed on an \gls{iot} device it may serve a variety of an attacker's goals such as to provide wrong application data to the \gls{iot} backend or to use the device as a bot in \gls{ddos} attacks, e.g., as in the Mirai attack \cite{mirai}. 

Malware can be detected through a security primitive called remote attestation, which has been excessively studied in the \gls{iot} domain (see \cite{Abera2016, Francillon2014, Steiner2016} for an overview), and standardized by the \gls{tcg} in the \gls{dice} series of specifications \cite{DICEhwReq, DICEattestation, DICEattestationSym, DICElayering}.
In a nutshell, attestation consists of 1) a method to securely calculate a fingerprint of a device's software stack and 2) a secure protocol conveying the fingerprint to a remote verifier that can evaluate the fingerprint. If the expected and the received fingerprints differ, the verifier may assume the device is running untrusted, potentially malicious firmware.

In contrast to \gls{tcg}'s series of attestation specifications using a dedicated security chip called \gls{tpm} \cite{tpm}, \gls{dice} is mainly intended to be used in so-called \textit{deeply embedded systems}. Deeply embedded systems are embedded systems relaying on constrained 8-, 16- or 32-bit microcontrollers running application software on top of a simple \gls{rtos} or bare metal (without an \gls{os}). 
Due to its low hardware requirements, \gls{dice} is more suitable than a \gls{tpm} for this class of devices.
In this paper, we concentrate on microcontroller-based deeply embedded systems. More powerful devices, e.g., running Linux or Windows are out of scope.

\gls{dice} is a hardware-software architecture supplementing the boot process of a microcontroller. The firmware a \gls{dice}-based device is separated into successively executed layers.
\gls{dice} provides measured boot through all firmware layers where in the top layer an attestation key depending on the measurements of all layers and a \gls{uds} is used in an attestation protocol with a backend.
If malware exists in some of the firmware layers the attestation key will be different and the protocol will fail. In this way, \gls{dice} can detect malware that is present on the device when the device boots.

In this paper, we demonstrate a \gls{toctou} attack on \gls{dice}-based attestation. We show that, \gls{dice} cannot detect malware installed at runtime which is capable of storing the valid attestation key in the flash memory of the microcontroller and reuse it across boot cycles for attestation instead of the newly derived key. We consider such malware capabilities as realistic and therefore we believe that attacks of this type are likely to happen in real \gls{iot} deployments.

Previous work in the area of \gls{toctou} attacks on attestation has either considered high-end \gls{tpm}-based Linux devices \cite{Bratus2008} or discuss \gls{toctou} countermeasures \cite{Nunes2020}. To the best of our knowledge we present the first attack on the standardized \gls{dice} attestation architecture for deeply embedded systems.

Our attack is conducted through \gls{rop} which is a control hijacking attack. Therefore, the requirements for our attack are the same as the requirements for \gls{rop} namely: 1) a memory corruption vulnerability and 2) attacker's knowledge of the devices' binary image. 

Memory corruption vulnerabilities are found often in embedded software as recently demonstrated in \cite{Kol2020, Senrio}. Such attacks are common for software written in system languages lacking memory safety such as C and C++ \cite{Szekeres2013}.  
Recent research \cite{Abbasi2019} has shown that even well-known protections such as
\gls{aslr}, 
\gls{dep}, and 
\gls{ssp} are almost completely absent in deeply embedded systems. 

An attacker can gain a copy of the firmware binary running on a fleet of \gls{iot} devices with the same software stack, e.g., by capturing or baying a single device and conducting a physical attack on its debug interfaces as shown, e.g., in \cite{Khan2020}. The firmware is required for identifying \gls{rop} gadgets, as well as for acquiring knowledge about the stack frame and the location of the attestation key in memory. Once the attacker has access to the binary she can develop the remote \gls{rop} attack described in this paper and use it against all devices belonging to a given fleet running the same software stack.

For our proof of concept implementation, we selected the Cortex-M architecture since Cortex-M is the defacto standard architecture used by all major semiconductor vendors for low-cost \gls{iot} \glspl{soc} and microcontrollers. The differences between the various Cortex-M variants are insignificant for the described attack since we use a subset of instructions available on all of them.
In summary, our contributions are:
\begin{itemize}
    \item We show a remote attack procedure abstracted from any specific IoT end-device capable to undermine \gls{dice}-based attestation. Once \gls{dice} is circumvented the attacker can reuse the same technique to install useful for him malware, e.g., such using the device for \gls{ddos} attacks.
    \item We provide a detailed proof of concept implementation of the attack on a Cortex-M4 based nRF42840 \gls{ble} microcontroller using an artificial memory corruption bug (a buffer overflow). We demonstrate the remote nature of the attack by conducting it in a realistic IPv6 over \gls{ble} \gls{iot} network \cite{rfc7668}.
    \item We propose practical countermeasures such as firmware updates, secure boot, and introducing additional inputs in the attestation key derivation.
\end{itemize}

\section{Background}
This section provides background information required for understanding the rest of the paper. Additionally, we provide discussion of the related work regarding \gls{iot} attestation and \gls{toctou} attacks in \Cref{sec:related_work}. 
\subsection{Control Hijacking Attacks and Countermeasures for Deeply Embedded Systems}
In this paper, we use \gls{rop} for conducting our attack which is a form of a control hijacking attack. In the following, we review the state of the art of control hijacking attacks and the most common countermeasures against them with a focus on deeply embedded systems. Our goal is to show that these attacks are and will remain a serious threat for deeply embedded systems.

Control hijacking attacks on deeply embedded systems were shown for a variety of microcontroller architectures. One of the earliest works in this area is \cite{Francillon2008} where a \gls{rop} attack is used to install persistent malware in the memory of an Atmel's AVR microcontroller. In \cite{Pastrana2016} the same architecture was attacked again with \gls{rop}. However, here the authors consider the attack in a broader context demonstrating a worm capable to propagate to neighbor notes in the network. The installation of persistent malware through \gls{rop} was demonstrated as well on Cortex-M devices \cite{Weidler2017}. 
\gls{rop} attacks were presented as well on the currently gaining momentum in the microcontroller market RISC-V architecture in \cite{Jaloyan2020, Gu2020}. 

Deeply embedded systems run most often software written in C and C++ either without an \gls{os} (bare metal) or on top of \gls{rtos}. 
Such setups almost always lack virtual memory as well as memory separation between tasks and separation between user- and kernel space, as shown in a recent survey conducted with 42 embedded \glspl{os} \cite{Abbasi2019}. 
The lack of those features allows memory corruption bugs to become a root for exploitable vulnerabilities. 

Many countermeasures were developed against exploits permitted by memory corruption bugs. Some of the most known are, e.g., \gls{cfi} \cite{Abadi2009}, shadow stack \cite{Burow2019}, \gls{aslr} \cite{aslr}, \gls{dep} \cite{DEPMicrosoft} and \gls{ssp} \cite{ssp}.

\gls{cfi} seeks to restrict control-flow transitions in a program to the set of strictly required transitions for the program's correct execution. Unfortunately, \gls{cfi} relies on process isolation and fine-grained memory protection. Without these features, \gls{cfi} cannot provide any security guarantees \cite{Walls2019}. Even if recent research efforts \cite{Kawada2021, Walls2019, Bresch2020, Nyman2017} aim to provide \gls{cfi} for deeply embedded systems \gls{cfi} is not yet broadly adopted.  

Shadow stack techniques compare a protected copy of subroutines' return address against the return addresses on the stack \cite{Burow2019}. In this way manipulations of the return addresses, e.g., as used by \gls{rop} are detected. However, shadow stacks require some form of memory protection.   

\gls{aslr} is a technique that prevents attackers to use the same gadgets on different instantiation of the same program containing the same bug. This is achieved by randomizing the offsets of program segments (e.g. code and data). \gls{aslr} is not suitable for microcontrollers because it requires virtual memory and an \gls{mmu} \cite{Abbasi2019}.

\gls{dep} counters code injection attacks by making data memory not executable and code memory not writable by using a \gls{mpu} or an \gls{mmu}. However, \gls{dep} cannot stop control flow attacks such as \gls{rop}. Moreover, \gls{dep} is infeasible for many microcontrollers lacking even a simple \gls{mpu}.  

\gls{ssp} detects buffer overflows by placing a random value (called cookie or canary) between the return address and local buffers on the stack. Before a function returns the value of the cookies is checked in order to determine if a buffer has overflowed and change it. \gls{ssp} provides relatively weak protection since it only can detect a small subset of special errors namely return address pointers valuations \cite{Szekeres2013} and therefore cannot stop code reuse attacks such as \gls{rop}.

A recent survey \cite{Abbasi2019} presents an evolution of the adoption of \gls{aslr}, \gls{dep}, and \gls{ssp} in 42 embedded \glspl{os}, where 20 of the \glspl{os} are specially intended for deeply embedded systems. The authors conclude that those features are broadly available only on high-end \glspl{os}. 
From all 20 deeply embedded  \glspl{os} three support \gls{dep}, one supports \gls{ssp}, and non supports \gls{aslr}.

\subsection{ARM Cortex-M Processors}
In this section, we introduce some details of the Cortex-M architecture that are relevant for our attack. Cortex-M CPUs have 13 general-purpose registers (\texttt{r0 - r12}), a stack pointer register (\texttt{sp}), a link register (\texttt{lr}) and the program counter register (\texttt{pc}) \cite{Yiu2013}.  Further, every Cortex-M processor also has special registers, which contain information about the processor status and define operation states and interrupt/exception masking. 
A special register in all Cortex-M variants is the PRIMASK register. It is used for exception and interrupt masking. It is 1-bit wide and blocks all exceptions and interrupts, when set. Special registers are not memory mapped and can only be accessed with certain special register access instructions. Instructions to modify the PRIMASK value are \texttt{cpsie i} (sets PRIMASK) for enabling exceptions and interrupts and \texttt{cpsid i} (clears PRIMASK) for disabling them \cite{Yiu2013}.

A key memory section for the processor to operate is the stack memory. It is used for temporary storage of register data, local variables, and function parameters. To store and retrieve data from the stack, ARM processors provide the \texttt{push} and \texttt{pop} instruction. 
The current stack pointer is incremented (\texttt{pop}) and decremented (\texttt{push}) automatically after each execution of these instructions. This non-intuitive increment and decrement of the stack pointer is done because the stack grows from a high memory address (usually the top of the SRAM region) to a lower address \cite{Yiu2013}.

\subsection{Return Oriented Programming}
\gls{rop} attacks use short code snippets that are already present in the code and link them in an order allowing arbitrary programs to be executed. These short code snippets are called gadgets. 
Gadgets consist of a small instruction sequence ending with a return instruction. The return instruction is used to chain multiple gadgets together. Each gadget performs a specific computation task (load, store, arithmetic operations, etc.), so that the combination of gadgets creates arbitrary programs which the attacker can execute. To perform \gls{rop}, the attacker exploits a buffer overflow in order to place addresses and other values on the stack memory. These addresses point to certain gadgets, which can use other stack-placed values as parameters for computation \cite{Weidler2017}.

\begin{figure}[t]
	\centering
	\begin{tikzpicture}
	
    	\node[anchor=north west,align=left,minimum width=2cm,minimum height=1cm] (main1) at (-1,0) {\small \textit{Caller routine}\\\scriptsize \texttt{...}\\\scriptsize \texttt{bl subroutine}};
	\node[anchor=north west,align=left,minimum width=1cm,minimum height=.2cm] (sub1) at (3,0) {\small \textit{Subroutine}};
	\node[anchor=north west,align=left,minimum width=2cm,minimum height=1cm] (sub) at (3,-1.5) {\scriptsize \texttt{/*save registers, incl. lr*/}\\\scriptsize \texttt{push \{r4, r5, r6, lr\}}\\\scriptsize \texttt{...}\\\scriptsize \texttt{/*execute tasks*/}\\\scriptsize \texttt{/*(r4-r6, lr might be changed)*/}\\\scriptsize \texttt{...}\\\scriptsize \texttt{pop \{r4, r5, r6, pc\}}\\\scriptsize \texttt{/*restore registers and return*/}};
	\node[anchor=north west,align=left,minimum width=2cm,minimum height=1cm] (main2) at (-1,-5) {\scriptsize \texttt{/*back to caller routine*/}\\\scriptsize \texttt{...}\\\scriptsize \texttt{/*next instructions*/}};

	\draw[->,draw] (main1.south east) -- (sub.north west);
	\draw[->,draw] (sub.south west) -- (main2.north);

	\end{tikzpicture}
	\caption{Use of push and pop instructions}
	\label{fig:pushPop}
\end{figure}
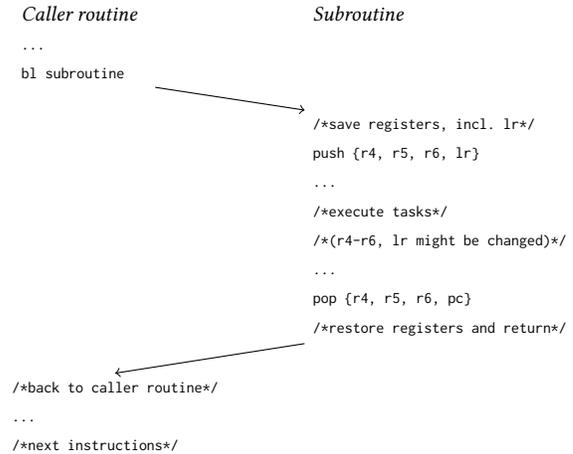

In general, ARM Cortex-M systems use push, pop, and branch instructions for control flow mechanisms. The \texttt{push} instruction is used by subroutines to store register values on the stack, including the return address held in the link register. When returning, either a branch instruction, e.g., \texttt{bx} or a \texttt{pop} instruction is executed.
A \texttt{pop} instruction is usually called at the end of subroutines to restore previously pushed registers including the return address from the stack. \Cref{fig:pushPop} shows this procedure. 
In \Cref{fig:pushPop} the return address is popped into the program counter register, which causes the processor to continue with the program code specified by the return address. This mechanism allows an attacker to exploit \texttt{pop} instructions to return to the address of a certain gadget, which ideally also includes a \texttt{pop} instruction in order to return to the next gadget. This assumes that, by buffer overflow, the attacker has previously placed appropriate address values on the stack \cite{Weidler2017}.

\subsection{Implicit Identity Based Device Attestation with DICE}\label{sec:dice}
Several variants of \gls{dice} attestation exist \cite{DICEattestation,DICEattestationSym, DICElayering}, where the differences are mainly in the number of software layers, the key derivations, and attestation protocols. Our attack is conducted on the top firmware layer and therefore it is applicable to all of them.
In the top firmware layer commonly resides the application logic of the device.
We base the following descriptions on the specification \textit{Implicit Identity Based Device Attestation} \cite{DICEattestation}.
This architecture consists of three layers --- the boot layer placed in ROM, the RIOT layer responsible for attestation key and certificate derivation, and the application layer. 
The RIOT layer receives a secret depending on its measurement and \gls{uds}. 
From this secret, it derives deterministically two asymmetric key pairs, one that is exclusive to the RIOT layer and depends only on the secret called DeviceID and one depending on the secret and the measurement of the application layer called alias key. The alias public key is certified with the DeviceID. Only the alias key and certificate are provided to the application layer for use in an attestation protocol with the backend.

\section{Assumptions}
For our attack we make the following assumptions:
\paragraph{The attacker can gain access to the firmware image.} We assume that the attacker can gain access to the firmware image, e.g., by conducting a physical attack on a captured (e.g., bought) device. For this, an image saved in external flash can be dumped or the read-back protection of the microcontroller can be circumvented. Attacks of the latter type are feasible since they were often demonstrated in the past, e.g., through glitching the debug interfaces \cite{LPC11attack, STM8attack} or using UV-light \cite{Obermaier2017}. A comprehensive summary of attacks targeting debug interfaces is given in \cite{Khan2020}. 
\paragraph{A fleet of \gls{iot} devices runs the same firmware.}
We consider a typical \gls{iot} scenario where many physically distributed  \gls{dice}-enabled devices run the same firmware stack. Once the attacker has developed and tested the attack on one captured device he can remotely conduct the attack on all devices of the fleet that run the same firmware. 
The goal of the attacker is to infect all devices running the same firmware and not the captured device.
\paragraph{The firmware contains an exploitable memory corruption bug.}
The firmware needs to provide an exploitable memory corruption bug, e.g., a buffer overflow vulnerability, which can be detected, e.g., by using techniques outlined in \cite{Rawat2012, Senrio}. Note that such vulnerabilities are common for embedded firmware written in system languages such as C/C++ \cite{Szekeres2013}, and were also often demonstrated in the past, e.g, \cite{Kol2020,Senrio}.
\paragraph{The firmware contains the necessary gadgets.}
The firmware has to contain suitable gadgets for the \gls{rop} attack. However, this is not a limiting requirement since our attack requires only two very simple gadgets. 
\paragraph{The attacker can gain knowledge of the stack frame structure.}
For conducting a \gls{rop} attack the attacker needs to gain knowledge about the stack frames. This information can be retrieved by flashing a captured device with the dumped image and using a debugger to examine the stack when a memory corruption is triggered.
\paragraph{The attacker can gain knowledge of the credentials' locations in memory.}
The memory locations of the attestation key and certificate need to be determined. This can be done by dumping the RAM when the top firmware layer is executed and finding the areas with high entropy.

\paragraph{Some functionality of the original firmware are still useful for the attacker.}
We assume also that parts of the original firmware, e.g.,
the network stack, the attestation protocol, etc. are still useful for the
attacker.

\section{Attack Method}\label{sec:attack_method}
The goal of an attacker is to persist code on the devices' flash memory that serves his purposes, e.g., code that will turn the devices of a given fleet  into bots executing \gls{ddos} attacks. This code, henceforth called \textit{useful malware}, should stay undetected by \gls{dice} attestation. In order to circumvent \gls{dice} we additionally install a small \textit{utility malware}. Both the useful and the utility malware are installed by exploiting a memory corruption bug through \gls{rop}. The main idea is that the utility malware  saves ones persistently the valid attestation credentials and then causes at every new reboot that they are used for attestation with the backend.

We install the utility malware first before installing the useful malware. It consists of two routines \texttt{ram2flash\_copy()} and \texttt{flash2ram\_copy()}. \texttt{ram2flash\_copy()} is executed only one single time just after installing the utility malware. It copies the valid alias private key and alias certificate from RAM to flash. After that, \texttt{ram2flash\_copy()} modifies the original firmware, e.g., some initialization function in the original firmware in order to cause an execution jump to \texttt{flash2ram\_copy()} every time the system boots. Then, at every subsequent boot, the \texttt{flash2ram\_copy()} overwrites the freshly calculated alias key and alias certificate with the previously saved old ones and jumps back to the regular firmware execution. This hides the modification of the device's firmware to any future attestation requests by the \gls{dice} backend. After that, the attacker exploits the same vulnerability for the second time in order to install the useful malware.

We divide our attack into five consecutive steps: 1) disabling interrupts, 2) installing utility malware to flash memory, 3) \texttt{ram2flash\_copy()} execution, 4) \texttt{flash2ram\_copy()} execution, and 5) installing useful malware to flash memory. These steps are explained in the following considering ideal gadgets. In \Cref{sec:implementation} we demonstrate how the same functionality is achieved with real gadgets easily found in firmware images.

\paragraph{Step 1: Disabling interrupts.}
As the attack is performed on a real-time microcontroller, first all interrupts need to be disabled. This prevents the \gls{rop} procedure and afterward the utility malware execution from being interrupted and experiencing unexpected behavior. To disable interrupts a gadget is used. This gadget needs to globally disable interrupts by means of a \texttt{cpsid i} instruction and afterward pop a new address value from the stack into the program counter to jump to the next gadget. An ideal gadget of this type is shown in \Cref{lst:gadget1_ideal}. To execute this gadget at the beginning of the \gls{rop} attack, the attacker has to replace the return address of the original stack frame with the 32-bit address of the gadget, followed by the address of the next gadget.

\begin{lstlisting}[caption=Ideal gadget to disable global interrupts (interrupt gadget), label=lst:gadget1_ideal, captionpos=b]
cpsid i
pop {pc}
\end{lstlisting}

\paragraph{Step 2: Installing utility malware to flash memory.}
To write to memory (and registers) with \gls{rop}, one single store gadget is sufficient. An ideal gadget of this type is shown in \Cref{lst:gadget2_ideal}. The first line stores a 32-bit value contained in register \texttt{rA} to a memory address contained in register \texttt{rB}. The second line gets new values for registers \texttt{rA}, \texttt{rB} and \texttt{pc} from the stack for the next execution of the store operation. Before the first execution of this gadget, execution of only the second gadget line is necessary to load the first values into the registers.

\begin{lstlisting}[caption=Ideal gadget to write values to memory (store gadget), label=lst:gadget2_ideal, captionpos=b]
str rA, [rB, #0]
pop {rA, rB, pc} 
\end{lstlisting}

To install the utility malware to flash memory the malware binary has to be split up into pieces of 32 bits. A 32-bit malware piece, a flash destination address, and the address of the first gadget line build an exploit data block of 96 bits, which is placed on the stack through a buffer overflow. The second gadget line pops the 32-bit malware pieces into register \texttt{rA}, the flash memory address into register \texttt{rB}, and the gadget address into the program counter register. This causes the next gadget call to store the malware pieces to the specified flash memory address.

After writing the utility malware to the flash memory, it is necessary to jump to \texttt{ram2flash\_copy()} to start executing it. Therefore, as the last part of the binary exploit, two 32-bit dummy values (pop into \texttt{rA} and \texttt{rB}) and the address of \texttt{ram2flash\_copy()} (pop into \texttt{pc}) have to be appended.

\begin{figure}[t]
	\centering
	\includegraphics[width=1\linewidth]{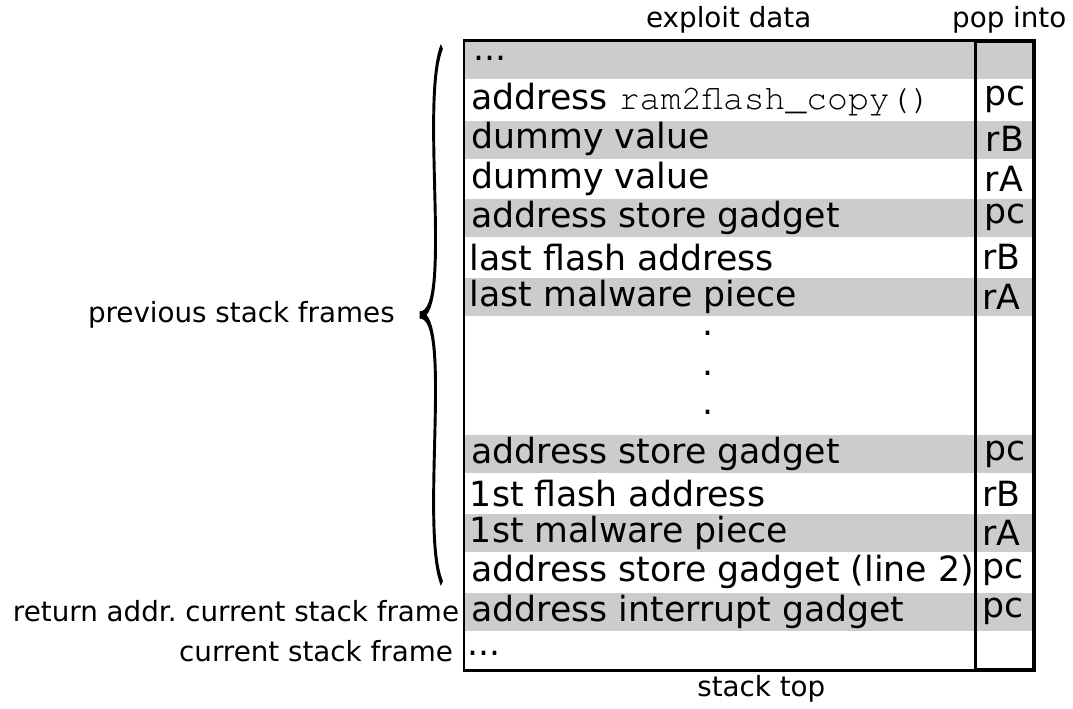}
	\caption{Stack memory after buffer overflow}
	\label{fig:stackExploit}
\end{figure}

\Cref{fig:stackExploit} shows the stack content required for installing the utility malware when the ideal gadgets from \Cref{lst:gadget1_ideal} and \Cref{lst:gadget2_ideal} are used. The left column shows which elements in the stack are overwritten by the exploit data through the buffer overflow. The middle column shows the exploit data. The right column shows in which registers the exploit data is popped during the \gls{rop} procedure. 

Note that usually, flash memory of a microcontroller needs to be configured before being able to write to it. As this is a device-specific task, this is not considered here. It will be explained in \Cref{sec:implementation}.

\paragraph{Step 3: \texttt{ram2flash\_copy()} execution.}
The function \texttt{ram2flash\_copy()} first needs to copy the private alias key and the alias certificate from RAM to flash memory.
This procedure preserves the current values in order to be used for future attestations. \Cref{lst:copyKeyCert} shows an example for that part of \texttt{ram2flash\_copy()} where key and certificate are assumed to be adjacent for simplicity.
\begin{lstlisting}[caption=Pseudo code of \texttt{ram2flash\_copy()} for coping alias key and certificate, label=lst:copyKeyCert, captionpos=b, float,floatplacement=H]
rA = key_cert_ram_address
rB = key_cert_flash_address
for j = 1 to (len(key) + len(cert)) do
	ldr rC, [rA, #0]
	str rC, [rB, #0]
	add rA, #4
	add rB, #4
end for
\end{lstlisting}
\texttt{ram2flash\_copy()} loads the RAM start address of the memory block holding key and certificate into register \texttt{rA} and the flash destination address into register \texttt{rB}. Then, it loads the word from the RAM address in register \texttt{rA} into register \texttt{rC}. Afterward, it stores the word to the flash address in register \texttt{rB}. To get the next addresses, it increments both addresses in registers \texttt{rA} and \texttt{rB} by four bytes. These steps are repeated until the combined length of the private key and certificate is reached and both key and certificate have been completely copied.

Once key and certificate are copied \texttt{ram2flash\_copy()} modifies the original firmware such that it jumps to \texttt{flash2ram\_copy()} at every system boot. It is important that the modified original firmware calls \texttt{flash2ram\_copy()} after the generation of the new attestation key and certificate, thus they get overwritten. A suitable firmware function to modify can be some initialization function, which is called on every application start. When modifying such a function, it is required that the modified function's functionality remains unchanged. Otherwise, arbitrary unexpected behavior might occur.

After modifying the original firmware the device needs to be reset or reinitialized. This is required because due to the buffer overflow the stack memory is corrupted and the regular operation of the device cannot continue. 
A reset can be triggered on many microcontroller architectures by writing into a special soft-reset register. For example, the Cortex-M devices provide Interrupt and Reset Control Register (AIRCR) for that purpose. Alternatively, a jump to the RAM initialization routine in the startup code can reinitialize the device.

\paragraph{Step 4: \texttt{flash2ram\_copy()} execution.}
After reset/re-initialization, the modified original firmware jumps to \texttt{flash2ram\_copy()}. \texttt{flash2ram\_copy()} takes the saved values for alias private key and alias certificate from flash memory and overwrites the freshly calculated values in RAM. 
For this, the routine from \Cref{lst:copyKeyCert} can be used again. It is only necessary to swap the address values of registers \texttt{rA} and \texttt{rB} in the beginning.
Additionally, \texttt{flash2ram\_copy()} has to jump back to the original program flow of the firmware. For this, it has to be made sure that potential return values of the original function are returned. To jump back, the \texttt{flash2ram\_copy()} can for example use a \texttt{bx lr} instruction, as the link register, which contains the return address, does not get changed during overwriting the key and certificate. Another possibility would be, in case the link register was pushed to the stack by the modified initialization function, to use a \texttt{pop {pc}} instruction.

\paragraph{Step 5: Installing useful malware.}
Now that the \gls{dice} attestation mechanism is bypassed, the attacker can again load malware to the device's flash memory undetected by the attestation process. For this, the attacker exploits the same buffer overflow as before. He again uses \gls{rop} to disables interrupts, writes the malware to the flash, modifies the original firmware at a suitable location to be able to jump to the useful malware, and resets the device. If the malware needs a large amount of memory which may not fit in the stack memory, the attacker can split up the malware and repeat this procedure multiple times.

\section{Proof of Concept Implementation}\label{sec:implementation}
In this section, we describe how we implemented the attack on our target device.

\subsection{Target Device - nRF52840}\label{sec:nrf}
We used the nRF52840 \gls{ble} microcontroller as a target for the attack. It is based on a 64\,MHz ARM Cortex-M4 processor. It has 1\,MB of flash memory and 256\,KB RAM. The flash memory is divided in 256 pages of 4\,KB each. 

To protect the flash memory from non-authorized access nRF52840 features an \gls{acl}. The \gls{acl} assigns and enforces access permissions to different regions of the on-chip flash memory map. The \gls{acl} protection can be activated at runtime and remains active until the next reboot.
In our \gls{dice} implementation we use the \gls{acl} to protect the \gls{uds} from writes and reads after it was used by the boot layer. This is done by configuring the \gls{acl} configuration registers during the boot layer execution.  

Further, the nRF52840 features a \gls{nvmc}. The \gls{nvmc} is used for writing and erasing the internal flash memory. Before writing to a flash page it has to be erased in advance, or it has to be empty. Also, flash erases can only be done for a whole page of 4\,KB at once. To erase a flash page the \gls{nvmc} configuration register has to be set to “erase enable”. Afterward, the starting address of the page has to be written to the \gls {nvmc} erasepage register, which starts the erase operation. To write values to addresses within the erased page the \gls{nvmc} configuration register has to be set to “write enable”. After this, values can be written to the flash page with a normal store instruction \cite{nrf52840}.

The radio software for the nRF52840 is provided within the \gls{sdk} \cite{sdk} as a pre-built software stack called softdevice \cite{s140}. We used the softdevice as a source for gadgets.

\subsection{Experimental Setup}
Our experimental setup consists of a nRF52840 evaluation board communicating with a laptop using IPv6 over \gls{ble} \cite{rfc7668}. The nRF52840 runs a UDP server application on top of an underlying \gls{dice} implementation. The UDP server application is taken from the nrf5 \gls{sdk} and uses the softdevice stack as \gls{ble} driver. On the laptop, a simple Python UDP client is implemented. 

We implemented \gls{dice} as explained in \Cref{sec:dice}. We placed the \gls{uds} into a flash page at address 0x000FF000 and protected it with the described \gls{acl}. Additionally, we use the \gls{acl} as well to protect the boot layer from overwriting.

To be able to perform the attack, a memory corruption vulnerability is required. In our proof of concept implementation, we use an artificial buffer overflow which was implemented inside a callback function that handles incoming UDP data. For this, we use a \textit{memcpy()} function that copies incoming data into a buffer on the stack without checking data length and buffer size.

\subsection{Identification of Gadgets}
We used the open-source tool ROPgadget \cite{ropgadget} to identify the required \gls{rop} gadgets within the softdevice \gls{ble} stack. For disabling interrupts a suitable gadget was found at memory address 0x00015EEE, see \Cref{lst:gadget1_found}. It is almost identical to the ideal gadget in \Cref{lst:gadget1_ideal}. The only difference is an additional register \texttt{r4} used as a parameter in the \texttt{pop} instruction. Fortunately, this additional register does not alter the function of the gadget, so it can be considered redundant. This means that an arbitrary dummy value can be put on the stack and popped into register \texttt{r4} when the gadget is used.

\begin{lstlisting}[caption=Gadget found to disable global interrupts (interrupt gadget), label=lst:gadget1_found, captionpos=b]
cpsid i                 ;cpsid i  (ideal) 
pop {r4, pc}            ;pop {pc} (ideal) 
\end{lstlisting}

A code section that fits the previously described store gadget in \Cref{lst:gadget2_ideal} can be found at memory address 0x00002976, see \Cref{lst:gadget2_found}. Comparing the ideal gadget with the actual gadget, \texttt{rA} corresponds to \texttt{r5} and \texttt{rB} corresponds to \texttt{r4}, but with a different order in the \texttt{pop} instruction. The different order in the \texttt{pop} instruction is not a problem since we only need to swap the malware data word and the destination address in the exploit data placed on the stack. Moreover, this combination of a store and pop instruction is only available with an additional register \texttt{r6} as a parameter to the \texttt{pop} instruction. Fortunately, similar to the other gadget, \texttt{r6} can be loaded with arbitrary dummy values which do not change the intended purpose of the gadget.

\begin{lstlisting}[caption=Gadget found to write values to memory (store gadget), label=lst:gadget2_found, captionpos=b, float]
str r5, [r4, #0]        ;str rA, [rB, #0] (ideal)  
pop {r4, r5, r6, pc}    ;pop {rA, rB, pc} (ideal)
\end{lstlisting}
\vspace{-1em}

\subsection{Assembling the Exploit}
Due to the slightly different gadgets, adaptions are necessary for the \gls{rop} procedure described in \Cref{sec:attack_method}. To fill the additional registers of each pop instruction with dummy values, such values have to be inserted into the exploit data. More precisely, the dummy values have to be inserted in front of every gadget address. This is because the redundant registers, as shown in \Cref{lst:gadget1_found} and \Cref{lst:gadget2_found}, are right before the program counter in the register order of the \texttt{pop} instructions. 

To be able to write to a certain flash memory address the attacker has to configure the \gls{nvmc} as explained in \Cref{sec:nrf}. In our case, the flash memory is much larger than the firmware size of the regular application, so it is sufficient to just enable flash writing and store the malware to the empty flash sections. If this is not the case, the attacker needs to choose a certain flash page that is not critical for his intentions and erase it first. The \gls{nvmc} configuration value (write enable), the \gls{nvmc} configuration register's address, and the store gadget address have to be inserted into the exploit in front of the first malware piece. This configures the \gls{nvmc} before the first write to flash.

\begin{figure}[t]
	\centering
	\includegraphics[width=.95\linewidth]{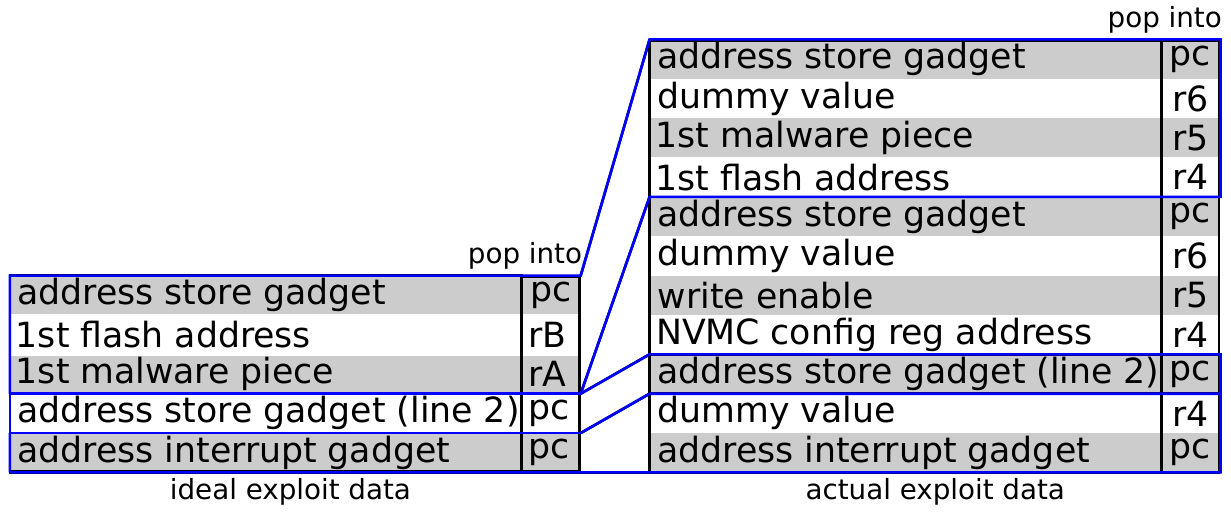}
	\caption{Comparison of ideal and actual exploit structure}
	\label{fig:ideal_vs_actual_exploid}
\end{figure}

\Cref{fig:ideal_vs_actual_exploid} shows how the flash configuration and the non-ideal gadgets affect the exploit data for our implementation compared to the ideal exploit.

The utility malware has to modify a function in the original firmware so that it jumps to \texttt{flash2ram\_copy()}, which copies the alias key and certificate from flash to RAM. As it is not possible to modify a flash memory that is not empty, the utility malware copies the whole flash page, containing the function to modify, into RAM, then modifies it, erases the original flash page, and writes back the modified page from RAM into flash. For this, a similar routine as shown in \Cref{lst:copyKeyCert} is used. The binary encodings of the new instructions are stored in flash as constant 32-bit words as part of the utility malware. The utility malware takes these words and replaces the original instructions at the right position inside the copied flash page in RAM. Afterward, after erasing the original flash page, again a slightly adapted routine as in \Cref{lst:copyKeyCert} is used to write back the modified flash page.

The two routines of the utility malware \texttt{flash2ram\_copy()} and \texttt{ram2flash\_copy()} are written in two distinct flash pages for simplicity of implementation.
\texttt{ram2flash\_copy()} is written to the flash page starting at memory address 0x000FC000. 
\texttt{flash2ram\_copy()} is written to the flash page at 0x000FE000. 
The flash page at the address 0x000FD000 is used for storing the alias key and certificate. 

We modified the original firmware function \texttt{app\_sched\_init()} in order to jump to \texttt{flash2ram\_copy()} at every boot. 
This function is part of the UDP server application and is located in the flash page starting at 0x00034000. The function initializes an event scheduler and is executed at every system start after \gls{dice} key derivation. We replaced the last 10 bytes of binary code in the function with the binary encoding of the instructions shown in \Cref{lst:jump_malware}, which also equals 10 bytes. This does not do any harm to the function's intended functionality. The first two instructions write the address of \texttt{flash2ram\_copy()} into register \texttt{r4}. Afterward, the CPU jumps to this address with a \texttt{bx} instruction. 

\begin{lstlisting}[caption=Instructions that replaced firmware code to jump to malware, label=lst:jump_malware, captionpos=b, float]
movw r4, #0xE000
movt r4, #0x000F
bx r4
\end{lstlisting}

\subsection{Analysis}
To evaluate our attack we implemented a \gls{dice} backend in Python on top of the Python UDP client. This backend executes a challenge-response protocol with the device, where the device signs a nonce with the alias private key. The signature and the alias certificate are sent to the backend. The backend verifies the signature of the nonce and the certificate. 

After sending the exploit data the device resets itself as intended. After reconnecting and sending an attestation request the attestation is still successful, even though we could clearly see, through inspecting the memory with a debugger, that the firmware was altered and that formerly empty flash pages are now filled with the utility malware, alias private key and alias certificate. This now allows the attacker to install useful malware without getting detected.

\Cref{tab:evaluation} gives an overview of the size of the \gls{rop} exploit and malware. The size of the \gls{rop} exploit is approximately four times the size of both utility malware routines combined. This is because, for every 32-bit malware piece, three additional 32-bit values are needed to build the exploit data and perform the \gls{rop} procedure. If an attacker wants to install complex useful malware that requires, e.g., about 100\,KB of memory, the \gls{rop} exploit would be about 400\,KB of size. Depending on how much stack memory is available or how much data the IPv6/UDP server implementation can handle at once, an attacker might need to split the useful malware into smaller parts. In our case, the limitation was the \gls{mtu} size of the IPv6 standard, which is 1,280 bytes. This means that in our case we would need to split a exploit of 400\,KB into approximately 375 to 400 parts. 
\begin{table}[h]
  \centering
  \begin{tabular}{ll}
    \toprule
    \textbf{Utility malware routine / \gls{rop} exploit data} & \textbf{Size} \\
    \midrule
    \texttt{ram2flash\_copy()} & 233\\
    \texttt{flash2ram\_copy()} & 52\\
    \gls{rop} exploit  & 1140\\
    \bottomrule
  \end{tabular}
  \caption{Sizes of utility malware routines and \gls{rop} exploit in bytes}
  \label{tab:evaluation}
\end{table}
 
\section{Countermeasures}
In this section, we discuss several approaches for mitigating the \gls{toctou} attack on \gls{dice} attestation and discuss their advantages and disadvantages. 
Additionally, we express our recommendation about the use cases in which a given countermeasure is a good/bad fit. 
%

\paragraph{Firmware Updates.}
If the vulnerability, allowing our attack becomes known to the device manufacturer, a firmware update can be provided. After this update, a new attestation key will be created and since the bug is removed our attack will not be possible. If the device refuses to attest with the new key, the device is still running malware. This means that \gls{dice} can detect malware with absolute certainty only if the bug allowing the malware becomes known and a patch is installed fixing it.
Note that providing updates that do not fix the vulnerability is insufficient since the attacker may compromise the newly installed firmware and get the new key by simply repeating the attack after each firmware update.
A drawback of this approach is that the bug may stay undetected for a long time.
Moreover, even if the bug becomes known, patching the vulnerability may not be possible because of diverse economic and technical reasons. 
This countermeasure can be preferred in use cases where the device manufacturer provides bug-fixing support for the devices over their lifetime and where it is acceptable that the devices run malware for the time needed to provide the patch. For use cases where this does not apply firmware updates are not a practical countermeasure.
\paragraph{Secure Boot with Secure Reset Trigger.}
One possible countermeasure against the described attack is to implement secure boot in parallel to \gls{dice} attestation. In contrast to \gls{dice} where the execution always jumps to the next layer regardless if the layer is compromised or not in the secure boot approach the jump in the next layer is accomplished only if the signature of the next layer is correct. Secure boot can restrict the effect of malware infection up to the time when the device resets.
However, if the device runs malware the malware may ensure that the device never resets. In order to overcome this problem, an additional mechanism is required allowing the backend to enforce a device reset as proposed in \cite{Huber2020, Xu2019}. In these papers, the authors propose the usage of an \gls{awdt} which causes the device to reset if the backend stops issuing authenticated tokens. However, this approach has higher hardware  requirements -- either an additional coprocessor \cite{Xu2019} or a \gls{tee}, e.g., TrustZone-M \cite{Huber2020}. Additionally, a reset may be disturbing in real applications. 
This countermeasure can be preferred in applications which can be safely reset. This countermeasure is especially not suitable for safety-critical applications, e.g., automotive control units where a reset of the CPU may cause dangerous situations. 
\paragraph{Additional Inputs in the Key Derivation Process.}
The key derivation process for the first mutable layer may use an additional input unique for the boot cycle. This input can be a nonce received from the backend before the last reset or a counter. Doing so will ensure that the attestation key is unique after each reboot. Note that this is a stronger approach than the secure boot approach (without reset trigger) since the backend can request attestation with a new nonce/counter which can only be provided if the device resets. In this way, a device running malware will not be capable to attest at any point in time.
This countermeasure can be preferred in the same use cases as the countermeasure using secure boot with secure reset trigger. However, its  advantage is that it does not require an additional chip or \gls{tee} and that the time of reset can be chosen by the application.
\paragraph{No Alias Key Exposure.}
Another possible countermeasure against our attack is to run the attestation protocol with the backend as the very first operation of the top firmware layer. Then the alias key must be deleted. This way the device does not expose any networking services and eventual vulnerabilities while the sensitive alias key is available. Additionally, the attestation protocol must be initiated by the device and must not require the receiving of any information from the backend such as a nonce. For replay protection of the protocol in this case a counter can be used, which however requires non-volatile storage. An additional  disadvantage is that for a new attestation, a reboot is required and this may be problematic for some real-world applications.
This countermeasure is equivalent to the countermeasure using additional inputs in the derivation process regarding the use case for which it is most suited.
%
%
\paragraph{Improving Memory Safety.}
The memory safety of the embedded software can be improved which will make it possible to avoid vulnerabilities such as buffer overflows. This can be achieved by using memory save languages such as Rust. In contrast to other memory save languages which may have too high requirements for the majority of microcontrollers, Rust is a compiled language that makes it comparable regarding speed of execution and executable size to C and C++. A disadvantage of this approach is that a large body of embedded C/C++ code already exists which needs to be rewritten. 
Using Rust can be beneficial in projects where the dependencies on existing code written in C and C++ are small. 

The advantages and disadvantages of the different techniques are summarized in \Cref{tab:countrameasures}. 
\begin{table*}[h]
    \centering
    \small
    \begin{tabular}{p{30mm}|p{50mm}|p{80mm}}
        \toprule
        \textbf{Method} & \textbf{Advantages} & \textbf{Disadvantages} \\ 
        \midrule
        firmware updates       & \textcolor{green}{$\oplus$} no additional hardware requirements      & \textcolor{red}{$\ominus$} the vulnerability may stay undetected for a long time \newline \textcolor{red}{$\ominus$} a patch needs to be developed\\ 
        \midrule
        secure boot \& \gls{awdt}  & \textcolor{green}{$\oplus$} timely malware detection          & \textcolor{red}{$\ominus$} requires additional chip or \gls{tee} \newline \textcolor{red}{$\ominus$} new attestations require reset     \\ 
        \midrule
        additional inputs     & \textcolor{green}{$\oplus$} timely malware detection          & \textcolor{red}{$\ominus$} requires persistent storage for nonce/counter   \newline \textcolor{red}{$\ominus$} new attestations require reset  \\ 
        \midrule
        no key exposure       & \textcolor{green}{$\oplus$} timely malware detection   & \textcolor{red}{$\ominus$} requires persistent storage for nonce/counter   \newline \textcolor{red}{$\ominus$} new attestations require reset\\ 
        \midrule
        using Rust      & \textcolor{green}{$\oplus$} no additional hardware requirements  & \textcolor{red}{$\ominus$} requires rewriting existing code     \\ 
        \bottomrule
    \end{tabular}
    \caption{Comparison of the different countermeasures}
    \label{tab:countrameasures}
\end{table*} 
To use additional inputs in the key derivation process or reducing the key exposure appears to offer good level of security and have acceptable disadvantages for the majority of use cases.

\section{Related Work}\label{sec:related_work}
\paragraph{Attestation.}
Attestation techniques are typically classified into three groups regarding their hardware requirements -- \textit{hardware-based}, \textit{software-based} and \textit{hybrid}. 
Hardware-based techniques, e.g., \cite{tpm, Raj2015, Johnson2016, Anati2013} use either a dedicated security chip (\gls{tpm} \cite{tpm}) or on-chip trusted execution environment such as TrustZone \cite{trustZone} or SGX \cite{McKeen2013}. Hardware-based techniques are considered to cumbersome for deeply embedded systems.
Software-based techniques, e.g., \cite{Seshadri2005, Seshadri2004, Ankergard2021} do not have any specific hardware requirements. They leverage information about the required time for certain computations such as the calculation of checksums. Those techniques are applicable only when the communication channel between prover and verifier has constant delays, therefore they are not suitable for devices communicating over the Internet. 
Hybrid techniques, e.g., \cite{Hristozov2018,Eldefrawy2012, Koeberl2014, Brasser2015, Shepherd2021} have lower hardware cost than the hardware techniques and at the same time does not impose timing requirements on the communication as the software-based techniques which makes them the preferred choice for constrained \gls{iot} devices. These techniques usually use some form of deeply integrated hardware extensions of the \gls{cpu}.
Further, regarding the time at which the attestation evidence is generated the attestation techniques can be classified in \textit{attestation with boot time evidence generation} \cite{Schulz2017, England2016, DICEhwReq}, \textit{attestation with on-request evidence generation} \cite{Hristozov2018,Eldefrawy2012, Koeberl2014, Brasser2015, Shepherd2021} and \textit{attestation with self-initiated evidence generation} \cite{Carpent2018, Carpent2017, Ibrahim2017}.

\paragraph{\gls{toctou} Attacks on Attestation.}
A discussion of the \gls{toctou} problem in the context of on-request hybrid attestation for constrained devices is provided in \cite{Nunes2020}. As a solution, the authors propose a method that uses secure logging of the time of memory modification. Alternatively, the methods presented in \cite{Carpent2018, Carpent2017,Ibrahim2017} propose self-initiated measurements with periodic or unpredictable schedule.
Another possible approach to mitigate the \gls{toctou} attack is presented in \cite{Abera2016cflat}, where the control flow path of the software is attested. 
An investigation of the problem of memory consistency during an attestation measurement is provided in \cite{Carpent2018a}. The lack of memory consistency may allow \gls{toctou} attacks. As a solution, the authors propose different memory locking approaches.
A \gls{toctou} attack on \gls{tpm} attestation is demonstrated in \cite{Bratus2008}. The authors show that attestation can succeed when loaded critical code and data are modified after they are measured with a \gls{tpm}. For this attack a Linux kernel vulnerability that allows the attacker to manipulate the page table is required. For demonstration purposes, the authors developed a malicious kernel module that executes the attack. 
As a solution, a method is proposed using the \gls{mmu} and the \gls{tpm}, where protected trap handlers update the \gls{tpm} registers when write to a loaded and previously measured memory occurs.

\paragraph{\gls{dice}}
The \gls{dice} series of specifications \cite{DICEhwReq, DICEattestation, DICEattestationSym, DICElayering} allows a variety of implementations. Implementations using standard microcontrollers \cite{Jager2017,Hristozov2018}, as well as, hardware implementations \cite{Jager2020} were demonstrated by previous research. Moreover, \gls{dice} is available in commercial products such as the Microchip's CEC1702 \cite{microchipDICE} and the NXP's LPC5500 \cite{microchipNXP}. An architecture combining \gls{dice} and secure boot was recently formally verified in \cite{Tao2021}.

\section{Conclusion}
The \gls{dice} series of specifications provide a standardized mechanism for detecting malware running on constrained \gls{iot} devices. However, \gls{dice} does not protect from malware capable of copying valid attestation keys and reusing them while the device is compromised. This is, however, a very common malware capability, therefore such attack vectors are realistic.
Our attack requires an exploitable memory corruption bug, e.g., a buffer overflow and knowledge of the firmware running on the device. While the former is a valid assumption on embedded systems written in C/C++, the latter information can be obtained by capturing a single device and dumping its software. Then our attack can be conducted on all \gls{iot} devices belonging to a given fleet running the same firmware, therefore our attack is scalable.
Through our proof of concept implementation we demonstrated the feasibility of the attack in realistic \gls{iot} deployments, for instance, where devices communicate over IPv6. We proposed several countermeasures to mitigate the shortcomings of the \gls{dice} series of specifications.

\bibliographystyle{ACM-Reference-Format}
\bibliography{bibliography}

\end{document}